\documentclass[journal]{new-aiaa}
\usepackage[utf8]{inputenc}
\usepackage{textcomp}

\usepackage{graphicx}
\usepackage{amsmath}
\usepackage[version=4]{mhchem}
\usepackage{siunitx}
\usepackage{longtable,tabularx}
\usepackage{dsfont}
\usepackage{amsmath}
\usepackage{amsfonts}
\usepackage{algorithm}
\usepackage{algpseudocode}
\usepackage{booktabs}
\usepackage{multirow}
\usepackage{graphicx}

\usepackage{mathtools}

\newcommand{\vect}[1]{\boldsymbol{#1}}
\newcommand{\set}[1]{\mathcal{#1}}

\title{Interception-Driven Inverse Reachability for Engagement Zone Construction}

\author{Grant Stagg\footnote{PhD Candidate, Electrical and Computer Engineering, Brigham Young University.}}
\author{Cameron K. Peterson\footnote{Associate Professor, Electrical and Computer Engineering, Brigham Young University, and AIAA Senior Member.}}
\affil{Brigham Young University, Provo, Utah, 84602, USA}

\author{Alexander Von Moll\footnote{Aerospace Engineer, Control Science Center, Air Force Research Laboratory, AIAA Member.}}
\author{Isaac Weintraub\footnote{Electronics Engineer, Control Science Center, Air Force Research Laboratory, AIAA Associate Fellow.}}
\affil{Air Force Research Laboratory, WPAFB, Ohio, 45433, USA}

\begin{document}

\maketitle


\begin{abstract}
In contested environments, autonomous vehicles may need to plan around adversarial pursuers whose launch locations are unknown. This paper presents an interception-driven inverse-reachability framework for inferring a feasible pursuer launch region directly from observed interception events for a single pursuer. Each interception induces a geometric constraint on the unknown launch location, and intersecting these constraints yields a bounded set guaranteed to contain the true origin under maximum-capability assumptions. Mapping this inferred set through the pursuer reachable region produces deterministic engagement zones with an explicit worst-case safety interpretation. A probabilistic extension models uncertainty in the pursuer launch location and yields graded engagement-risk fields for risk-aware planning. To accelerate localization, we introduce an information-driven planner for sacrificial agents that selects trajectories to maximize expected contraction of the feasible launch region. Monte Carlo simulations show that the proposed framework rapidly reduces launch-location uncertainty and enables substantially shorter safe trajectories after only a small number of sacrificial deployments.
\end{abstract}
\let\thefootnote\relax\footnotetext{Distribution Statement A. Approved for public release: distribution is unlimited. AFRL-2026-1807. Cleared 13 Apr 2026.}

\section{Introduction}
Autonomous vehicles operating in contested environments must reason about adversarial threats whose launch locations are often unknown~\cite{chapman2025safe}. Traditional engagement analysis typically assumes that pursuer parameters—such as launch position, range, and maneuverability—are available \emph{a priori} or described by a known probability distribution~\cite{von2023basic,stagg2025acc}. In many operational settings, however, threat locations must instead be inferred from indirect, sparse, or event-driven observations, motivating active localization methods that infer hidden sources from sparse binary contacts or geometry-dependent measurements~\cite{abraham2018data,hoffmann2023non}. This creates a coupled estimation--planning problem in which agents must gather informative data about a threat while simultaneously generating trajectories that remain safe during information acquisition~\cite{lew2022safe}.

Interception events provide direct geometric information about the pursuer origin. If a pursuer successfully intercepts an agent, it must have launched from a location capable of reaching the interception location within its travel budget. Each interception therefore imposes a spatial constraint on the unknown launch location. Intersecting these constraints yields a feasible launch region that is guaranteed to contain the true origin under bounded-capability assumptions. As additional interceptions are observed, this region contracts, progressively reducing uncertainty about the threat location.

Crucially, a bounded launch region induces a bounded reachable region (RR), the
set of positions that a pursuer originating anywhere within the feasible launch
region could attain. Expressed in the evader frame, this RR defines a
corresponding engagement zone (EZ) that contains all evader positions from
which interception could occur by a pursuer launched from any admissible
location. Any trajectory whose position remains outside the EZ corresponding
to the evader's instantaneous heading is safe against all admissible pursuers.
Thus, interception-driven localization enables safety-critical planning without
requiring precise knowledge of the adversary's launch location.

Despite this connection, existing EZ formulations fundamentally rely on prior knowledge of pursuer launch locations or assume that a probabilistic uncertainty model is available beforehand~\cite{von2023basic,chapman2025engagement, stagg2025acc,stagg2026probabilistic}. In many operational environments these assumptions are unrealistic, as threat parameters must instead be inferred from indirect and event-driven observations. Comparatively little attention has been given to constructing EZs directly from interception data.

This paper eliminates the need to know the pursuer’s launch location in advance. It introduces a geometric framework that uses interception events to infer the feasible launch region and then propagates this region forward to construct both deterministic and probabilistic EZs.
Each interception generates a set constraint on the unknown launch location; under assumptions that the pursuer has a bounded range, these constraints take the form of discs whose intersection defines a region guaranteed to contain the true origin. 
Using the inferred launch region, we compute all positions the pursuer could possibly reach. This produces a worst-case EZ that guarantees safety outside its boundary, while a probabilistic extension instead measures how likely interception is based on the amount of geometric overlap.

Building on this inference structure, sacrificial agents are deployed as deliberate information-gathering assets. By inducing interceptions at strategically selected locations, these agents actively contract the feasible launch region, improving operational awareness and enabling progressively less conservative trajectory planning for high-value assets.

The primary contributions of this work are as follows:
\begin{itemize}
    \item A geometric inverse-reachability method for inferring a feasible pursuer launch region from interception events for a single pursuer.
    \item A reachable-region construction that maps bounded launch uncertainty into deterministic EZs, that can be used to plan safe trajectories.
    \item A probabilistic extension that quantifies engagement likelihood via geometric overlap, enabling risk-aware planning.
    \item An information-driven trajectory objective that selects sacrificial paths to maximize expected contraction of the feasible pursuer launch-region.
\end{itemize}

This paper proceeds as follows. Section~\ref{sec:related_works} reviews related work, and Section~\ref{sec:BEZ} provides background on EZs. Section~\ref{sec:gbez} examines EZs under bounded launch uncertainty, and Section~\ref{sec:gpez} extends this formulation to the probabilistic setting. Section~\ref{sec:ez_from_interceptions} presents methods for constructing EZs from interception events. Section~\ref{sec:safe_paths} and Section~\ref{sec:sacrificial_path} then use these regions for safe trajectory generation and informative sacrificial-agent path planning. Results are presented in Section~\ref{sec:results}, followed by conclusions in Section~\ref{sec:conclusion}.

\subsection{Related Work}\label{sec:related_works}

Engagement-zone (EZ) reasoning is rooted in pursuit--evasion theory and differential games
\cite{isaacs1965differential,weintraubPursuit}, where geometric interception conditions
establish the relationship between relative kinematics, aspect angle, and range constraints.
Modern work has focused on analytic EZ constructions that preserve geometric interpretability
while remaining tractable for planning and safety analysis. For simple motion kinematics, Basic engagement zones (BEZs)
are formalized in \cite{von2023basic}, with subsequent extensions incorporating turn-constrained pursuers in~\cite{chapman2025engagement}. Optimal-control formulations enable planners to treat engagement
zones as dynamic obstacles \cite{weintraub2022optimal}, while multi-threat settings motivate
trajectory optimization methods capable of handling densely contested environments
\cite{dillon2023optimal,milutinovic2025stochastic}. Sampling-based planners for large
collections of engagement constraints are presented in \cite{wolek2024sampling}, and
range-limited pursuer navigation with explicit safety guarantees is developed in
\cite{chapman2025safe}. Related threat-aware weaponeering analyses building on BEZ
structure are also emerging \cite{vonmoll2025weaponeering}.

Probabilistic engagement zones (PEZs) extend these ideas by replacing binary safe/unsafe
boundaries with a probability of interception under parameter uncertainty. Prior work
developed PEZ models for turn-constrained pursuers and integrated probabilistic constraints
into risk-aware planning \cite{stagg2025acc,stagg2026probabilistic}. Despite these advances,
existing engagement-zone formulations fundamentally assume that pursuer launch locations are
known or that uncertainty is characterized by a prescribed prior distribution. In many
operational settings, however, threat locations must instead be inferred from sparse,
event-driven observations. Constructing EZs directly from interception data
remains largely unexplored.

The inference component of this work is closely related to bounded-error localization and
set-membership estimation. Classical trilateration relies on equality constraints that can
yield point estimates given sufficiently rich geometry \cite{trilateration}. By contrast,
capability-based observations naturally induce inequality constraints, producing feasible
regions rather than isolated solutions. This perspective has been studied in sensor-network
localization and geometric constraint-based estimation
\cite{terwilliger2008localization,mourad2011robust,shi2017robust,lee2009localization,huang2017practical,singh2017geometric},
with extensions addressing measurement uncertainty in aerial systems
\cite{sorbelli2022measurement} and related modalities such as acoustic detection
\cite{jekaterynczuk2023survey}. In contrast to prior localization work that estimates a
source position, the present paper infers a feasible launch \emph{set} for a maneuver-constrained
pursuer and maps this set forward through reachability to construct EZs that
provide explicit safety guarantees.

Finally, the use of sacrificial agents to deliberately collect informative data appears in
multi-robot exploration under resource constraints \cite{cesare2015multi} and heterogeneous
uncertainty-aware team planning \cite{duggan2025uncertaintyawareplanningheterogeneousrobot},
as well as in prior radar-inference work \cite{stagg2025cooperative}. However, the present setting is
distinct: information is obtained only through interception outcomes, so trajectory design
acts as active feasible-set contraction within a threat-learning pipeline.

\subsection{Engagement Zones and Pursuit--Evasion Geometry}\label{sec:BEZ}
EZs describe regions in which a faster pursuer can intercept an evader given their relative kinematics and range constraints.
In the classical two-player pursuit--evasion game, both vehicles move at constant speeds and headings, with dynamics
\begin{equation}
\dot{\vect{x}}_P =
v_P
{\renewcommand{\arraystretch}{0.8}
\begin{bmatrix}
\cos\psi_P \\
\sin\psi_P
\end{bmatrix}
},
\quad
\dot{\vect{x}}_E =
v_E
{\renewcommand{\arraystretch}{0.8}
\begin{bmatrix}
\cos\psi_E \\
\sin\psi_E
\end{bmatrix}
},
\label{eq:ez_dynamics}
\end{equation}
where $\vect{x}_P,\vect{x}_E \in \mathbb{R}^2$ denote the pursuer and evader positions, and $v_P,v_E$ their respective speeds.
The pursuer is assumed to have a finite range $R$ and to capture the evader once the distance between them becomes less than a capture radius $r$.
The speed ratio $\nu = v_E/v_P < 1$ specifies the evader's relative speed.

\begin{figure}[ht]
    \centering
\includegraphics{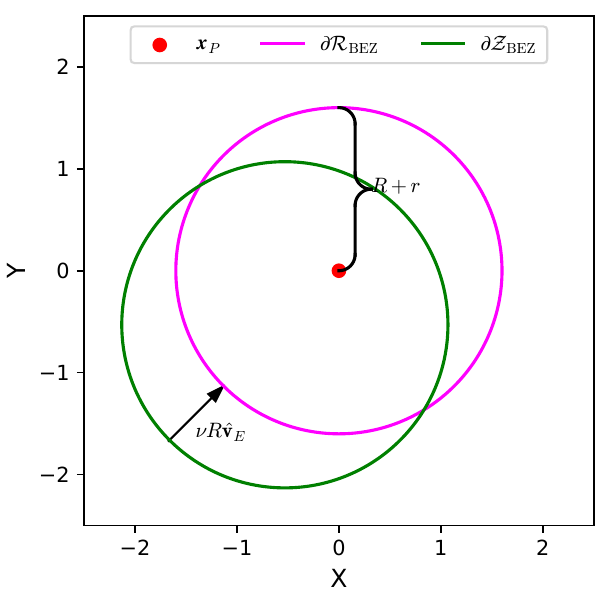}
\caption{Pursuer position (red), RR (magenta), and BEZ (green) are shown for an evader heading of \ang{45}.} \label{fig:BEZ}
\end{figure}

\emph{Basic engagement zones} (BEZs) are defined as the set of initial evader positions from which interception can occur before the pursuer exhausts its range.
For a pursuer located at the origin, this region can be expressed as~\cite{von2023basic}
\begin{equation}
\set{Z} =
\left\{
\vect{x}_E :
\|\vect{x}_E - \vect{x}_P(0)\|_2
\le
\rho(\xi,\nu,R,r)
\right\},
\label{eq:bez_definition}
\end{equation}
where $\rho(\xi,\nu,R,r)$ specifies the boundary distance as a function of the aspect angle $\xi$ between the evader's heading and the line of sight to the pursuer.
The boundary distance function is found geometrically~\cite{von2023basic};
\begin{equation} \label{eq:rho}
    \rho(\xi, \nu, R,r) = \nu R\left(\cos(\xi)+\sqrt{\cos^2(\xi)-1+\frac{(R+r)^2}{\nu^2R^2}}\right),
\end{equation}
where $\xi$ is the aspect angle between the agent's heading and the pursuer defined as
\begin{equation}
\xi(\vect{x}_A,\vect{x}_P(0)) = \psi_A - \operatorname{atan2}\left({y_P(0)-y_A},{x_P(0)-x_A}\right).
\end{equation}
Although this form provides analytical insight, it depends explicitly on the instantaneous aspect angle and does not generalize easily when the pursuer's location is uncertain or inferred.

The BEZ can be interpreted as a translated version of the pursuer's RR. The RR defines the set of all positions the pursuer can reach within its travel budget, including the capture radius, and therefore forms a disc of radius $R+r$ centered at the pursuer's position. The BEZ is obtained by translating this region by the distance the evader travels while the pursuer expends its full range, namely $\nu R$ in the direction opposite the evader's heading.

Figure~\ref{fig:BEZ} illustrates this construction for an evader heading of \ang{45}. The boundary of the RR is shown in magenta, the BEZ boundary in green, and the evader trajectory is indicated.

\section{Geometric Engagement Zones Under Launch Uncertainty}
This section develops a geometric formulation of EZs under
bounded uncertainty in the pursuer launch location. When the initial pursuer
position is unknown but constrained to lie within a set, the resulting RR is likewise bounded, producing a family of EZs that admit a natural
worst-case safety interpretation. These geometric constructions form the basis
for both deterministic EZs and the probabilistic engagement zones (PEZs)
introduced later in this section.

\subsection{Deterministic Engagement Zones}\label{sec:gbez}
We first consider the deterministic worst-case construction, in which the
pursuer may launch from any point in a feasible set.

\subsubsection{Geometric Interpretation Using the Minkowski Sum}

A geometrically transparent representation of EZs is obtained through set operations. Let $\set{P} \subset \mathbb{R}^2$ denote the feasible pursuer launch region. We define the RR as the set of all points that the pursuer can reach within its travel budget while accounting for the capture radius. This region is given by the Minkowski sum
\begin{equation}
\set{R} = \set{P} \oplus \mathcal{D}(\vect{0}, R + r),
\label{eq:RR}
\end{equation}
where $\mathcal{D}(\vect{x}_c,\rho) = \{\vect{x} \in \mathbb{R}^2 : \|\vect{x}-\vect{x}_c\|_2 \le \rho\}$ denotes a disc of radius $\rho$ centered at $\vect{x}_c$, and
\begin{equation}
\set{A} \oplus \set{B} = \{\vect{a} + \vect{b} : \vect{a}\in \set{A},\, \vect{b}\in \set{B}\}
\label{eq:minkowski_sum}
\end{equation}
denotes the Minkowski sum of two sets $\set{A},\set{B} \subseteq \mathbb{R}^2$.

Geometrically, this construction corresponds to a uniform offset of the launch region by distance $(R+r)$, thereby characterizing all points that are reachable—and therefore capturable—by the pursuer from some feasible launch location. 
If the launch region $\set{P}$ is convex, the RR $\set{R}$ is likewise convex, since Minkowski addition preserves convexity.


\subsubsection{Engagement Zone as a Shifted Reachable Region}

Because the evader continues to move during an encounter, the set of points that can lead to interception is displaced in the direction opposite the evader's motion.
Let $\hat{\vect{v}}_E = [\cos\psi_E,\,\sin\psi_E]^\top$ denote the unit vector along the evader's heading.
The EZ can then be defined as a translation of the pursuer's capture region:
\begin{equation}
\set{Z}
= \set{R} - \nu R \hat{\vect{v}}_E
= \left\{
\vect{x}_E \in \mathbb{R}^2 :
\vect{x}_E + \nu R \hat{\vect{v}}_E \in \set{R}
\right\}.
\label{eq:EZ_shift}
\end{equation}
In this form, a point $\vect{x}_E$ lies inside the EZ if, after translating it forward by the evader's possible displacement during the engagement ($\nu R$), it falls within the pursuer's capture region.
This shifted RR interpretation generalizes the constant-heading BEZ geometry to arbitrary or inferred pursuer regions $\set{P}$.

\begin{figure}[ht]
    \centering
\includegraphics{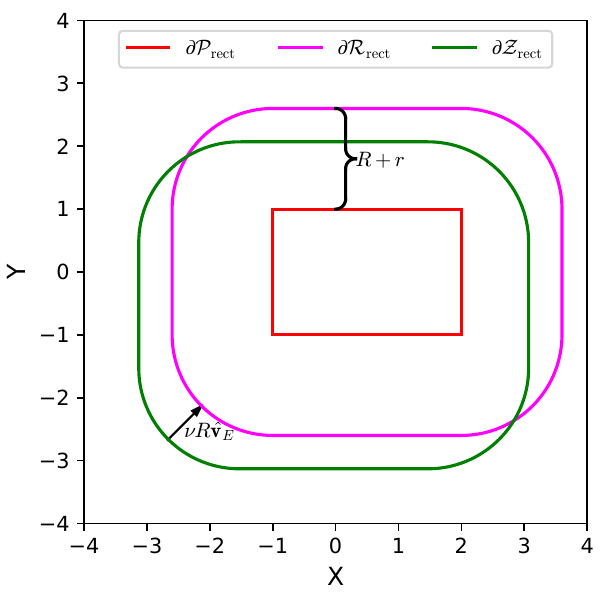}
\caption{The boundary of $\set{P}_{\textit{rect}}$ (red), $\set{R}_{\textit{rect}}$ (magenta), and $\set{Z}_{\textit{rect}}$ (green) are shown for an evader heading of \ang{45}.} \label{fig:rect_BEZ}
\end{figure}
As an example, consider the case in which the feasible pursuer launch region is a rectangular region defined by lower and upper bounds on the pursuer’s Cartesian coordinates, denoted $\vect{x}_P^{\textit{min}}$ and $\vect{x}_P^{\textit{max}}$. This set is written as
\begin{equation}
    \set{P}_{\textit{rect}}(\vect{x}_P^{\textit{min}},\vect{x}_P^{\textit{max}}) =
    \left\{
        \vect{x}_P \in \mathbb{R}^2 :
        \vect{x}_P^{\textit{min}} \le \vect{x}_P \le \vect{x}_P^{\textit{max}}
    \right\},
    \label{eq:rectangular_P}
\end{equation}
where the inequalities are interpreted elementwise.

The corresponding pursuer RR is obtained by taking the Minkowski sum of this rectangle with a disc of radius $(R + r)$,
\begin{equation}
    \set{R}_{\textit{rect}}
    = \set{P}_{\textit{rect}} \oplus \set{D}(\vect{0}, R + r),
    \label{eq:rectangular_R}
\end{equation}
which produces a rectangle inflated by $(R + r)$ and rounded by quarter-circle arcs at its corners.
Finally, the EZ is obtained by translating this region opposite the evader’s direction of motion by $\nu R$,
\begin{equation}
    \set{Z}_{\textit{rect}}
    = \set{R}_{\textit{rect}} - \nu R \hat{\vect{v}}_E
    = \{\vect{x}_E \in \mathbb{R}^2 :
    \vect{x}_E + \nu R \hat{\vect{v}}_E \in \set{R}_{\textit{rect}}\}.
    \label{eq:rectangular_EZ}
\end{equation}
Geometrically, this results in an EZ that resembles a rounded rectangle shifted
in the direction opposite the evader’s heading by a distance proportional to the
speed ratio $\nu$, as illustrated in Figure~\ref{fig:rect_BEZ}. In the figure,
the boundary of the feasible pursuer launch region $\partial \set{P}_{\textit{rect}}$
is shown in red. Dilating this set by $R+r$ produces the pursuer capture region
$\set{R}_{\textit{rect}}$, whose boundary $\partial \set{R}_{\textit{rect}}$ is shown
in magenta. Translating this region opposite the evader's heading (\ang{45} in
this case) yields the EZ $\set{Z}_{\textit{rect}}$, whose boundary
$\partial \set{Z}_{\textit{rect}}$ is shown in green. The figure also shows the
evader trajectory $\nu R \hat{\vect{v}}_E$, representing the path the evader
would follow if the pursuer were to expend its full range.

\subsection{Geometric Probabilistic Engagement Zones}\label{sec:gpez}

The previous section presented the worst-case EZ, defined under the assumption that the pursuer may launch from any point in the region $\set{P}$ and that the evader must remain safe against all such possibilities. If, instead, the evader can tolerate some level of risk, a probabilistic characterization is more appropriate. This section introduces the geometric probabilistic engagement zone (PEZ), which quantifies the probability that a pursuer drawn from a distribution of launch positions can reach the evader.

In this formulation, the pursuer’s launch position is modeled as a random variable with probability density function (PDF)
\begin{equation}
    \vect{x}_P\sim f_{\vect{x}_P}(\vect{x}).
\end{equation}
For any measurable region $\set{X} \subset \mathbb{R}^2$, the probability that the pursuer launches from within $\set{X}$ is given by integrating the PDF over that region:
\begin{equation}
    \mathbb{P}(\vect{x}_P \in \set{X})
    = \int_{\vect{x} \in \set{X}} f_{\vect{x}_P}(\vect{x}) \, d\vect{x}.
\end{equation}

\subsubsection{Probabilistic Reachability Under Launch Uncertainty}

The probability that an arbitrary point $\vect{x}$ is reachable by a pursuer whose launch location is uncertain is obtained by integrating over all possible launch positions. For the infinite--turn-rate geometric model, the pursuer can reach a point if its launch position lies within a disc of radius $(R+r)$ centered at $\vect{x}$. Thus, the reachability probability is
\begin{equation}\label{eq:RR_prob_def}
    p_{\mathrm{RR}}(\vect{x})
    = \mathbb{P}(\mathrm{pursuer\ can\ reach}\;\vect{x})
    = \int_{\vect{\tau} \in \mathbb{R}^2}
        \mathds{1}\!\left( \|\vect{\tau} - \vect{x}\| \le R + r \right)
        f_{\vect{x}_P}(\vect{\tau}) \, d\vect{\tau},
\end{equation}
where $\mathds{1}(\cdot)$ is the indicator function
\begin{equation}
    \mathds{1}(z) =
    \begin{dcases}
        1, & \text{if } z \text{ is true}, \\[-4pt]
        0, & \text{otherwise}.
    \end{dcases}
\end{equation}
This expression evaluates the probability mass of all pursuer launch positions lying inside the disc of radius $(R+r)$ centered at $\vect{x}$, and therefore quantifies the probability that a pursuer drawn from the distribution $f_{\vect{x}_P}$ can reach location $\vect{x}$.

\subsubsection{Probabilistic Reachability in the Evader Frame}\label{subsec:shift_prr}

As in the deterministic case, the evader’s motion during an engagement requires that reachability be evaluated at a shifted location. An evader at position $\vect{x}_E$ is effectively equivalent to the translated point
\begin{equation}
    \vect{x}_E^{\mathrm{shift}} = \vect{x}_E + \nu R \,\hat{\vect{v}}_E,
\end{equation}
which accounts for the evader’s displacement during the encounter. Thus, the probability that the pursuer can reach the evader from a randomly distributed launch position is obtained by evaluating the reachability probability at this shifted point:
\begin{equation}
    p_{\mathrm{EZ}}(\vect{x}_E)
    = p_{\mathrm{RR}}(\vect{x}_E + \nu R \,\hat{\vect{v}}_E).
\end{equation}
This transformation directly maps the probability of pursuer reachability, defined in the pursuer-fixed frame, to the evader’s starting location, thereby yielding the probability that the evader lies inside the PEZ.

\subsubsection{Uniform Rectangular Launch Distribution}

A simple and illustrative special case arises when the pursuer's launch location is assumed to be uniformly distributed over a rectangular region. Let the launch region be $\set{P}_{\textit{rect}}(\vect{x}_P^{\textit{min}},\vect{x}_P^{\textit{max}})$ as defined in Equation~\eqref{eq:rectangular_P},
with area
\begin{equation}
    A_P = (x_{\max} - x_{\min})(y_{\max} - y_{\min}).
\end{equation}
The pursuer's launch-position pdf is then
\begin{equation}\label{eq:rect_pdf}
    f_{\vect{x}_P}(\vect{x})
    =
    \begin{cases}
        \dfrac{1}{A_P}, & \text{if } \vect{x} \in \set{P}_{\textit{rect}}, \\[6pt]
        0,               & \text{otherwise}.
    \end{cases}
\end{equation}
Substituting this density into the reachability expression in \eqref{eq:RR_prob_def} yields
\begin{align}
    p_{\mathrm{RR}}(\vect{x})
    &= \int_{\vect{\tau} \in \mathbb{R}^2}
        \mathds{1}\!\left( \|\vect{\tau} - \vect{x}\| \le R + r \right)
        f_{\vect{x}_P}(\vect{\tau}) \, d\vect{\tau} \\
    &= \frac{1}{A_P} \int_{\vect{\tau} \in \set{P}_\textit{rect}}
        \mathds{1}\!\left( \|\vect{\tau} - \vect{x}\| \le R + r \right)
        d\vect{\tau}.
\end{align}
The indicator function restricts the integral to the portion of the rectangle $\set{P}_\textit{rect}$ that lies inside the disc of radius $(R+r)$ centered at $\vect{x}$. Thus,
\begin{equation}\label{eq:rect_prr}
    p_{\mathrm{RR}}(\vect{x})
    = \frac{
        \mathrm{Area}\!\left(
            \set{P}_{\textit{rect}} \,\cap\,
            \set{D}(\vect{x},R+r)
        \right)
    }{
        \mathrm{Area}(\set{P}_{\textit{rect}})
    },
\end{equation}
where $\mathrm{Area}(\cdot)$ denotes planar area.

In words, the reachability probability $p_{\mathrm{RR}}(\vect{x})$ is equal to the fraction of the rectangular launch region that lies within the capture disc centered at $\vect{x}$. Applying the shift to evader coordinates from the previous subsection, the PEZ value at an evader starting location $\vect{x}_E$ becomes
\begin{equation}
    p_{\mathrm{EZ}}(\vect{x}_E)
    = p_{\mathrm{RR}}(\vect{x}_E + \nu R \,\hat{\vect{v}}_E)
    = \frac{
        \mathrm{Area}\!\left(
            \set{P}_{\textit{rect}} \,\cap\,
            D\big(\vect{x}_E + \nu R \hat{\vect{v}}_E,R+r\big)
        \right)
    }{
        \mathrm{Area}(\set{P}_{\textit{rect}})
    }.
\end{equation}
This provides a simple geometric interpretation of the PEZ for a uniformly distributed pursuer launch position: at each evader location, the engagement probability is the normalized area of overlap between the launch rectangle and the shifted capture disc. 

\begin{figure}[ht]
    \centering
\includegraphics{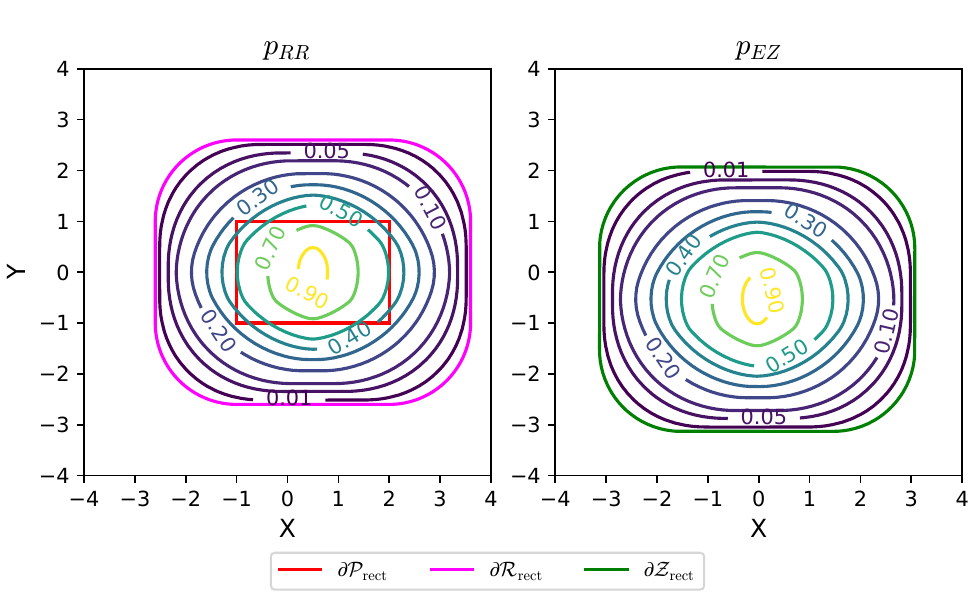}
\caption{The $p_{\textrm{RR}}$ (left), and $p_{\textrm{EZ}}$ (right) are shown. The boundary of $\set{P}_{\textit{rect}}$ (red), $\set{R}_{\textit{rect}}$ (magenta), and $\set{Z}_{\textit{rect}}$ (green) are shown for an evader heading of \ang{45}.} \label{fig:rect_PEZ}
\end{figure}

These probabilities are shown as level sets in Figure~\ref{fig:rect_PEZ}. 
The left plot illustrates $p_{\mathrm{RR}}$ together with the deterministic geometric 
boundary of the RR $\set{R}_{\textit{rect}}$. 
The right plot shows $p_{\mathrm{EZ}}$ for a pursuer with heading $45^{\circ}$, 
along with the corresponding deterministic EZ boundary. 
As the probability threshold decreases, both $p_{\mathrm{RR}}$ and $p_{\mathrm{EZ}}$ 
approach their respective deterministic boundaries. 
Indeed, the deterministic geometric boundaries coincide with the zero-probability 
contours of $p_{\mathrm{RR}}$ and $p_{\mathrm{EZ}}$.

\section{Inferring Engagement Zones from Interception Events}\label{sec:ez_from_interceptions}
In this section, we present a geometric method for constructing the potential EZs from observed interception locations of sacrificial evaders. This construction requires the pursuer’s maximum range and capture radius to be known, inferred from prior information, or conservatively assumed. We first develop an approach that assumes interception location is the only available measurement. We then extend the method to incorporate pursuer launch-time information when such data are available.

\subsection{Inference from Interception Location}\label{sec:inf_from_int}
Let $\{\vect{x}_{\textit{int}}^{(i)}\}_{i=1}^{N_{\textit{int}}}$ denote the set of locations at which sacrificial agents have been intercepted, where $N_{\textit{int}}$ is the total number of interceptions. For any single interception location $\vect{x}_{\textit{int}}^{(i)}$, the pursuer must have been capable of reaching that point within its travel budget. Consequently, the pursuer's launch position must lie within a disc of radius $R+r$ centered at the interception location. This defines the feasible launch region region as
\begin{equation}
\set{P}_{\textit{pot}}\left(\vect{x}_{\textit{int}}^{(i)}\right)
= \set{D}\!\left(\vect{x}_{\textit{int}}^{(i)}, R+r\right).
\end{equation}
When multiple interception locations are available, each observation induces a reachability constraint on the pursuer's launch position. The feasible launch region must therefore satisfy all such constraints simultaneously, yielding the intersection of discs centered at the respective interception locations with radius $R+r$:
\begin{equation}
\set{P}_{\textit{pot}}\left(\{\vect{x}_{\textit{int}}^{(i)}\}\right)
=
\bigcap_{i=1}^{N_{\textit{int}}}
\set{D}\!\left(\vect{x}_{\textit{int}}^{(i)}, R+r\right).
\end{equation}

This set contains all launch positions from which the pursuer could have reached every observed interception point within its allowable travel distance. Because each reachability constraint is convex, their intersection is likewise convex, and its boundary forms a closed curve composed entirely of circular arcs. 

With two interception events, the feasible region forms a lens bounded by symmetric arcs. With three or more events, the region becomes a convex circular-arc polygon whose boundary consists of alternating circle--circle intersection vertices and arc segments. Degenerate cases may also arise: the intersection reduces to a single disc when one disc is contained within all others, and becomes empty when the constraints are mutually inconsistent.

This circular-arc representation provides an exact geometric characterization of the feasible pursuer launch region and serves as the foundation for subsequent engagement-zone construction. An algorithm for computing this boundary is presented in Section~\ref{sec:bound}.

An example of $\set{P}_{\textit{pot}}$ is shown in Fig.~\ref{fig:pot_BEZ}(a). Interception locations are marked with red crosses, and the corresponding reachability discs are depicted with dotted red boundaries. The boundary of the resulting feasible pursuer region is shown in solid red.

\subsection{Inference with Launch-Time Information}
If evaders can additionally measure the pursuer's launch time (e.g., via acoustic or visual sensing), tighter bounds on the pursuer's launch position can be established. For the $i^{\textit{th}}$ interception, let $t_{\textit{launch}}^{(i)}$ denote the pursuer's launch time and $t_{\textit{int}}^{(i)}$ the interception time. The available travel time is therefore
\begin{equation}
t_{\textit{trav}}^{(i)} = t_{\textit{int}}^{(i)} - t_{\textit{launch}}^{(i)}.
\end{equation}
Assuming the pursuer travels at a known constant speed $v_P$, the maximum distance it could have traversed is
\begin{equation}
d_{\textit{trav}}^{(i)} = v_P\, t_{\textit{trav}}^{(i)}.
\end{equation}
Consequently, the launch position associated with the $i^{\textit{th}}$ engagement must lie within a distance $d_{\textit{trav}}^{(i)} + r$ of the interception location.

For an interception at $\vect{x}_{\textit{int}}^{(i)}$ with measured travel distance $d_{\textit{trav}}^{(i)}$, the corresponding feasible pursuer launch region is the disc
\begin{equation}
\set{P}_{\textit{pot}}\bigl(\vect{x}_{\textit{int}}^{(i)}\bigr)
= \set{D}\bigl(\vect{x}_{\textit{int}}^{(i)},\, d_{\textit{trav}}^{(i)}+r\bigr).
\end{equation}
When multiple interceptions are observed, each with an associated launch-time measurement, the feasible launch region must satisfy all refined reachability constraints simultaneously. The resulting region is therefore given by the intersection of discs with event-dependent radii:
\begin{equation}
\set{P}_{\textit{pot}}\bigl(\{\vect{x}_{\textit{int}}^{(i)}\}\bigr)
= \bigcap_{i=1}^{N_{\textit{int}}}
\set{D}\bigl(\vect{x}_{\textit{int}}^{(i)},\, d_{\textit{trav}}^{(i)}+r\bigr).
\end{equation}

As in the case where launch time is not measured, the feasible launch region
boundary is formed by arcs of circles. When launch time is measured, however,
these arcs have varying radii.
In general the boundary consists of circle--circle intersection vertices joined by arc segments of varying radii. An algorithm for computing this boundary is presented in Section~\ref{sec:bound}.

Figure~\ref{fig:pot_BEZ}(b) illustrates an example of $\set{P}_{\textit{pot}}$. Interception locations are marked with red crosses, and the associated reachability discs are shown with dotted red boundaries. The resulting feasible launch region is rendered in solid red. Notably, for the same set of interception locations, incorporating launch-time information produces a substantially smaller feasible region than the fixed-range construction.

\subsection{Feasible Launch-Region Boundary Construction}\label{sec:bound}
Because the feasible launch region is convex and bounded by circular arcs, its boundary can be computed exactly using pairwise disc intersections.
We compute the boundary $\partial \set{P}_{\textit{pot}}$ as a collection of circular arcs induced by the intersection of reachability discs. Each interception event generates a disc
\begin{equation}
    \set{D}^{(i)} = \set{D}\bigl(\vect{x}_{\textit{int}}^{(i)},\, \rho^{(i)}\bigr),
\end{equation}
where $\rho^{(i)} = R + r$ in the interception-only case and $\rho^{(i)} = d_{\textit{trav}}^{(i)} + r$ when launch-time information is available.

For every unordered pair of discs $(\set{D}^{(i)}, \set{D}^{(j)})$, we compute up to two circle--circle intersection points. A candidate intersection point $\vect{x}$ is retained only if it lies within all discs, i.e., $\vect{x}\in \set{D}^{(k)}$ for $k = 1, \dots, N_{\textit{int}}$. The resulting set of feasible vertices is deduplicated using a geometric tolerance $\varepsilon$.

For each disc $\set{D}^{(i)}$ with center $\vect{c}^{(i)} = \vect{x}_{\textit{int}}^{(i)}$ and radius $\rho^{(i)}$, the retained vertices on its circumference define angular coordinates
\begin{equation}
    \theta = \operatorname{atan2}(x_y - c_{i,y},\, x_x - c_{i,x}).
\end{equation}
Sorting these angles produces consecutive vertex pairs. For each pair $(\theta_a^{(i)},\theta_b^{(i)})$, the corresponding circular arc on $\set{D}^{(i)}$ is included if its midpoint lies within all discs, thereby ensuring that only true boundary segments of the intersection are retained.

The resulting boundary representation is given by the set of arcs
\begin{equation}
    \set{A}
    =
    \left\{
        \bigl(c^{(i)},\; \rho^{(i)},\; \theta^{(i)}_{\mathrm{start}},\;
        \theta^{(i)}_{\mathrm{end}}\bigr)
    \right\}_{i\in[1,\hdots,N_a]},
\end{equation}
where $N_a$ is the number of arcs which together define an ordered, piecewise-circular closed curve. Because the feasible launch region is convex, these arcs uniquely characterize its boundary. If one disc is strictly contained within all others, the procedure yields a single circle. If the discs have an empty intersection, no arcs are produced. The procedure is summarized in Algorithm~\ref{alg:bound}.
\begin{algorithm}[h]
\caption{Exact Boundary Extraction for the Intersection of Discs}
\label{alg:bound}
\begin{algorithmic}[1]
\State \textbf{Input:} Disc centers $\{c^{(i)}\}$, radii $\{\rho^{(i)}\}$, tolerance $\varepsilon$
\State \textbf{Output:} Boundary arc set $\set{A}$

\State Compute all pairwise circle--circle intersection points
\State Retain only points that lie within every disc (to tolerance $\varepsilon$)
\State Sort and deduplicate the remaining points to obtain the vertex set $\set{V}$

\For{each disc $\set{D}^{(i)}$ with center $c^{(i)}$ and radius $\rho^{(i)}$}
    \State Identify vertices in $\set{V}$ that lie on $\partial \set{D}^{(i)}$
    \State Sort these vertices by polar angle about $c^{(i)}$
    \For{each consecutive angle pair $(\theta_a,\theta_b)$}
        \State Compute midpoint angle $\theta_m = (\theta_a+\theta_b)/2$
        \State Compute midpoint $\vect{x}_m = c^{(i)} + \rho^{(i)}[\cos\theta_m,\sin\theta_m]$
        \If{$\vect{x}_m$ lies within every disc}
            \State Add arc $(c^{(i)}, \rho^{(i)}, \theta_a, \theta_b)$ to $\set{A}$
        \EndIf
    \EndFor
\EndFor

\State \Return $\set{A}$
\end{algorithmic}
\end{algorithm}

\subsection{Engagement Zones Induced by the Feasible Launch Region}
\begin{figure}[h]
    \centering
\includegraphics{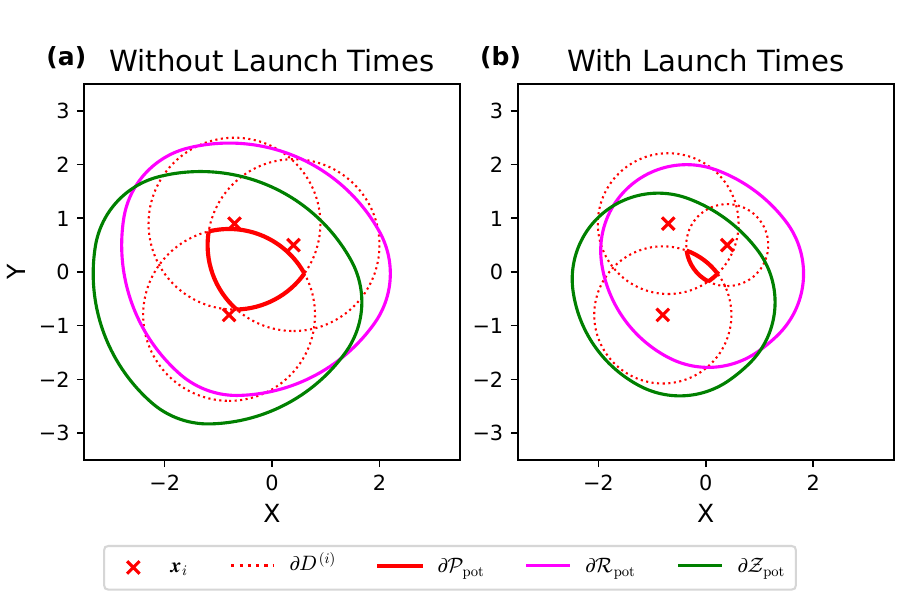}
\caption{Interception locations (red $\times$'s), discs (dotted red), pursuer region (red), RR (magenta), and EZ (green) for \ang{45}. (a) Without launch timing; (b) with launch timing.} \label{fig:pot_BEZ}
\end{figure}

With the feasible pursuer launch region defined, the corresponding EZ can be constructed. We first dilate the feasible launch region by a disc of radius $R+r$ to obtain the potential capture region
\begin{equation}
    \set{R}_{\textit{pot}}
    = \set{P}_{\textit{pot}} \oplus \set{D}\bigl(\vect{0}, R+r \bigr).
\end{equation}
This operation corresponds to the Minkowski sum of the feasible launch set with the pursuer’s reachability disc.

The dilation is implemented via a signed distance function (SDF) derived from the circular-arc boundary of $\set{P}_{\textit{pot}}$. Because the boundary consists exclusively of arc segments, the SDF reduces to the minimum Euclidean distance from a query point to the boundary, with the sign determined by set membership. Minkowski addition with a disc is equivalent to offsetting the SDF by $(R+r)$; accordingly, the dilated region is obtained by subtracting $(R+r)$ from the SDF of $\set{P}_{\textit{pot}}$.

The potential EZ is obtained by translating the capture region in the direction opposite the evader's heading:
\begin{equation}
    \set{Z}_{\textit{pot}}
    = \set{R}_{\textit{pot}} - \nu R \hat{\vect{v}}_E
    = \{\vect{x}_E \in \mathbb{R}^2 :
    \vect{x}_E + \nu R \hat{\vect{v}}_E \in \set{R}_{\textit{pot}}\}.
    \label{eq:potential_EZ}
\end{equation}
Geometrically, this translation accounts for the evader’s motion during the pursuer’s travel time, yielding the set of evader positions that remain vulnerable to interception.

Figure~\ref{fig:pot_BEZ} depicts the RR (magenta) and EZs (green) induced by the feasible pursuer launch region for both the launch-time case (a) and the no-timing case (b). Because launch-time information contracts the feasible launch region, the resulting RRs and EZs are correspondingly smaller and more tightly bounded.

\begin{figure}[h]
    \centering
\includegraphics[width=.9\linewidth]{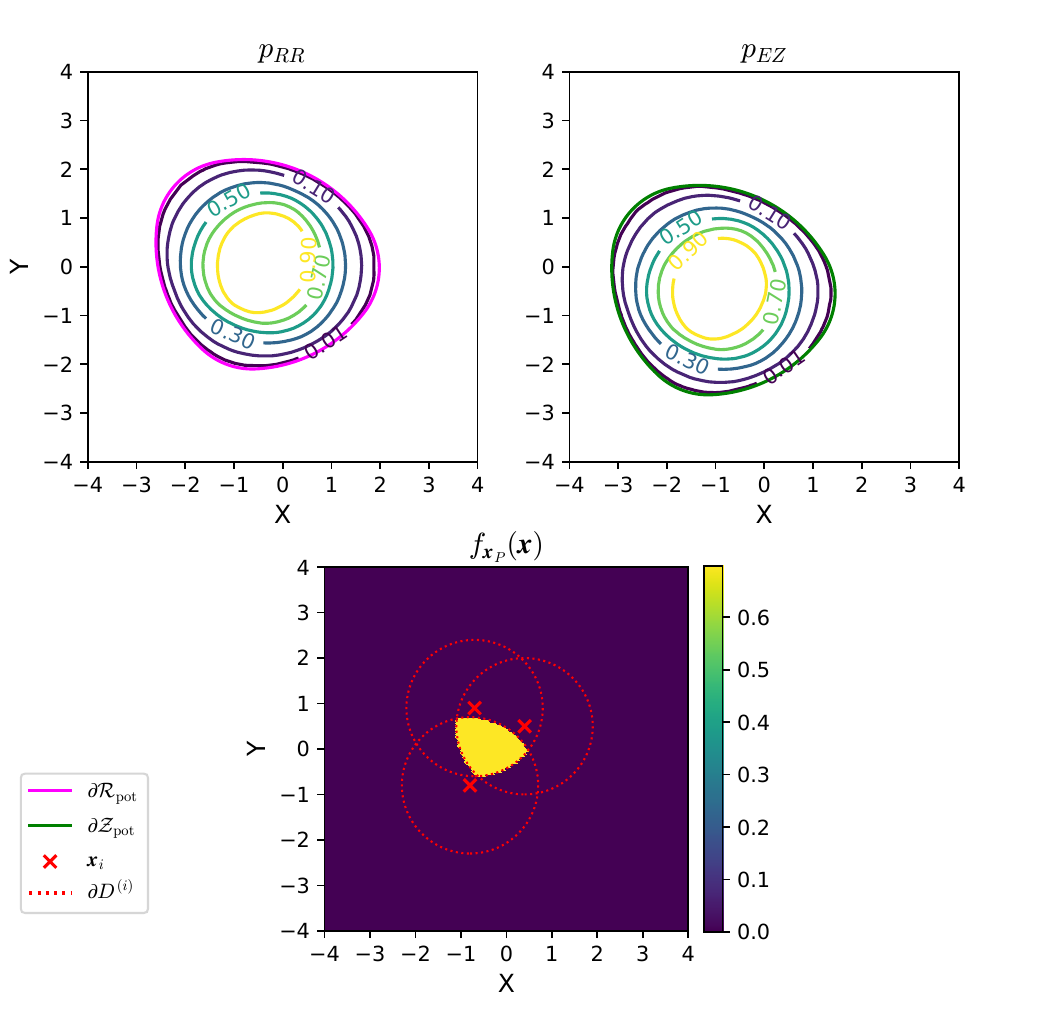}
\caption{The $p_{\textrm{RR}}$ (left), and $p_{\textrm{EZ}}$ (right) are shown. The boundary of $\set{R}_{\textit{pot}}$ (magenta) and $\set{Z}_{\textit{pot}}$ (green) are shown for an evader heading of \ang{45}. The lower figure shows the uniform PDF of the pursuer's launch location.} \label{fig:pot_PEZ}
\end{figure}

\subsection{Probabilistic Engagement Zones}
The reachability probability and the corresponding probabilistic engagement
zone (PEZ) were introduced in Section~\ref{sec:gpez}. For the infinite-turn-rate
geometric pursuer model, the reachable-region probability (RRP) at a point
$\vect{x}$ is the probability that the pursuer launch location lies within a disc
of radius $(R+r)$ centered at $\vect{x}$.

As shown in Section~\ref{sec:inf_from_int}, each interception induces a
disc constraint on the pursuer launch location, and the intersection of these
constraints defines the feasible launch region $\mathcal{P}_\textit{pot}$. In the
absence of additional prior information, we model the pursuer launch-location
density as uniform over $\mathcal{P}_\textit{pot}$:
\begin{equation}
f_{\vect{x}_P}(\boldsymbol{\tau})
=
\begin{cases}
\displaystyle \frac{1}{\mathrm{Area}(\mathcal{P}_\textit{pot})}, & \boldsymbol{\tau}\in\mathcal{P}_\textit{pot},\\[6pt]
0, & \text{otherwise}.
\end{cases}
\end{equation}

Under this maximum-entropy assumption, the RRP reduces to the geometric form
\begin{equation}\label{eq:pot_prr}
p_{\mathrm{RR}}(\vect{x})
=
\frac{
\mathrm{Area}\!\left(
\mathcal{P}_\textit{pot} \cap \mathcal{D}(\vect{x}, R+r)
\right)
}{
\mathrm{Area}(\mathcal{P}_\textit{pot})
}.
\end{equation}
Thus, the RRP is determined entirely by geometric overlap between the reachability disc and the feasible launch region.

Figure~\ref{fig:pot_PEZ} illustrates the RRP and the resulting PEZ. The upper-left panel displays level sets of the RRP together with the geometric RR shown in magenta. All probability level sets are fully contained within the RR, as expected from the underlying reachability constraint.
The upper-right panel depicts the PEZ alongside the geometric EZ for an evader heading of \ang{45}. The bottom panel presents the launch-location density, observed interception locations, and the circular feasibility constraints that define the pursuer launch region.
Notably, incorporating launch-region uncertainty smooths the sharp boundary of the geometric reachable set, producing a graded notion of engagement risk.

\section{Planning Under Inferred Engagement Constraints}
In this section, we present two path-planning methods that leverage geometric EZs and PEZs. The first method, presented in Section~\ref{sec:safe_paths}, generates safe trajectories for high-value, non-attritable agents by avoiding potential engagement with the pursuer. The second method, presented in Section~\ref{sec:sacrificial_path}, designs trajectories for sacrificial, attritable agents that act as information-gathering platforms to maximize information about the pursuer's unknown launch location.

\subsection{Trajectory Planning Under Engagement Constraints}\label{sec:safe_paths}
We first consider trajectory generation for high-value agents that must avoid
potential engagement. The planner enforces safety by imposing constraints derived
from geometric EZs and PEZs. 
In general, such constraints may be constructed from any assumed pursuer launch
region $\mathcal{P}$ and, in the probabilistic setting, an associated launch-location
PDF $f_{\vect{x}_P}(\vect{x})$. The emphasis of this work, however, is on planning safe
trajectories using EZs and PEZs derived directly from interception observations,
as developed in Section~\ref{sec:ez_from_interceptions}. This connection enables
threat-aware motion planning without requiring prior knowledge of the pursuer’s
state.

The evader trajectory is parameterized by a B-spline curve and optimized to
minimize travel time. Because the evader operates at constant speed, minimizing
final time is equivalent to minimizing path length. Kinematic feasibility is
maintained through bounds on curvature and turn rate, ensuring that the resulting
trajectory is physically realizable. Safety is enforced by constraining the path
to remain outside the EZ induced by the feasible pursuer launch region or by enforcing the maximum PEZ probability to be less than a safety threshold $\varepsilon$.

B-splines are selected for their smoothness properties, differentiability, and
local support, which together promote a sparse derivative structure and improve the efficiency of gradient-based solvers. Let the control points be
$\mathcal{C} = (\vect{c}_1, \vect{c}_2, \ldots, \vect{c}_{N_c})$, and define the
knot vector

\begin{equation}
\vect{t}_k =
\big(t_0 - k\Delta_t, \ldots, t_0 - \Delta_t,\,
t_0,\,
t_0 + \Delta_t, \ldots, t_f,\,
t_f + \Delta_t, \ldots, t_f + k\Delta_t\big),
\end{equation}
where $k$ denotes the spline order, $t_0$ and $t_f$ are the initial and final
times, $N_k$ is the number of interior knots, and
$\Delta_t = (t_f - t_0)/N_k$ is the uniform spacing. The trajectory is then
\begin{equation}
\label{eq:bspline}
\vect{p}(t) = \sum_{i=1}^{N_c} B_{i,k}(t)\vect{c}_i,
\end{equation}
where the basis functions $B_{i,k}(t)$ are constructed via the Cox--de Boor
recursion~\cite{cox1972numerical}.

The objective is to compute a feasible trajectory connecting the initial state
$\vect{x}_0$ to the goal $\vect{x}_f$ while minimizing the final time.
Feasibility is defined with respect to both vehicle kinematics and engagement
avoidance. Assuming a unicycle motion model with bounded curvature and turn
rate, the resulting nonlinear program is
\begin{subequations}
\label{eq:optimization}
\begin{align}
(\mathcal{C}_{\mathrm{opt}}, t_{f_{\mathrm{opt}}})
&= \arg\min_{\mathcal{C}, t_f} \; t_f, \\
\text{s.t.}\quad
&\vect{p}(0) = \vect{x}_0, \\
&\vect{p}(t_f) = \vect{x}_f, \\
&\vect{p}(t_s)\notin \mathcal{Z}, \qquad \forall t_s \in \mathcal{T}_s,\label{eq:ez_constraint}\\
&v_E(t_s) = v_E, \qquad \forall t_s \in \mathcal{T}_s, \\
&u_{\mathrm{lb}} \le u_E(t_s) \le u_{\mathrm{ub}}, \qquad \forall t_s \in \mathcal{T}_s, \\
&|\kappa_E(t_s)| \le \kappa_{\mathrm{ub}}, \qquad \forall t_s \in \mathcal{T}_s,
\end{align}
\end{subequations}
where $\mathcal{Z}$ is the EZ and
$\mathcal{T}_s = \{0,\Delta_t,2\Delta_t,\ldots,t_f\}$ is the set of discrete
collocation times with $\Delta_t = t_f/N_t$ at which the constraints are enforced.
The EZ constraint requires the evader to remain outside of the EZ generated from the feasible launch region. When using geometrically constructed EZs, this ensures that
no point along the trajectory is reachable by a pursuer originating within the
feasible launch region. To enable a tradeoff between safety and path optimality,
a PEZ constraint may be used instead. Replacing
\eqref{eq:ez_constraint}, one may enforce
\begin{equation}
    p_{\mathrm{EZ}}\!\left(\vect{p}(t_s)\right) \le \varepsilon, \qquad \forall t_s \in \mathcal{T}_s,
\end{equation}
where $p_{\mathrm{EZ}}(\cdot)$ denotes the probability that a state lies inside
the true EZ and $\varepsilon \in [0,1]$ is a user-specified risk threshold.

The remaining constraints enforce kinematic consistency.
Under the unicycle model, the kinematic feasibility constraints can be found as
\begin{equation}\label{eq:kinematic}
v_E(t) = \|\dot{\vect{p}}(t)\|_2,\qquad
u_E(t) =
\frac{\dot{\vect{p}}(t)\times\ddot{\vect{p}}(t)}
{\|\dot{\vect{p}}(t)\|_2^2},\qquad
\kappa_E(t) = \frac{u(t)}{v(t)},
\end{equation}
where $v$ is the velocity, $u$ is the turn rate, and $\kappa$ is the curvature of the trajectory. Since the evader speed is fixed at \(v_E\), the turn rate and curvature satisfy
\begin{equation}
    \kappa_E(t) = \frac{u_E(t)}{v_E}.
\end{equation}
Therefore, the constraints on \(u_E(t)\) and \(\kappa_E(t)\) are equivalent up to scaling by \(v_E\). In practice, it is sufficient to impose only one of them, using the value corresponding to whichever physical limitation is more restrictive.

The optimization problem is solved using the interior-point solver
IPOPT~\cite{ipopt}. Derivatives of both the objective and constraints are
computed through automatic differentiation in JAX~\cite{jax2018github},
allowing efficient construction of sparse Jacobians and Hessian
approximations suitable for large-scale nonlinear programs.

\subsection{Sacrificial Trajectories for Launch-Region Inference}
\label{sec:sacrificial_path}
Sacrificial trajectory design is formulated as an information-gathering
optimal control problem in which interception events serve as measurements
that contract the feasible pursuer launch region.
Trajectories are selected to deliberately induce informative
interceptions that accelerate geometric inference of the pursuer launch location.
In essence, sacrificial agents are converted into active sensing assets,
where the trajectory design directly governs the rate at which launch-location
uncertainty is reduced.
Because information is obtained only upon interception, trajectory selection
is built upon a probabilistic event model coupled with the geometric
consequences of an observed interception.

\subsubsection{Interception Hazard Model}

Let the sacrificial agent trajectory be parameterized by a B-spline
$\vect{p}(t)$ and sampled at times $\{t_k\}_{k=1}^{K}$,
producing points $\{\vect{p}(t_k)\}_{k=1}^{K}$.
At each position, we evaluate the probability that a pursuer drawn from the
current posterior over launch locations and range could reach the evader.

Interception is modeled as a terminal event that may occur at any point along
the trajectory using a discrete hazard formulation. Let \(\Delta t_k\) denote
the time increment associated with sample \(k\). We assume the interception
hazard is piecewise constant over each interval
\([t_k,\, t_k+\Delta t_k)\), so that risk accumulates with time spent in the
threat field. Specifically, the hazard over step \(k\) is defined as
\begin{equation}
    \lambda_k = \alpha\, p_{\mathrm{RR}}\bigl(\vect{p}(t_k)\bigr),
\end{equation}
where \(\alpha>0\) is an event-intensity parameter and
\(p_{\mathrm{RR}}(\vect{p}(t_k))\) is the probability that a pursuer can reach
the evader position \(\vect{p}(t_k)\), as defined in
Equation~\eqref{eq:RR_prob_def}.

Under this model, the probability of no
interception during the interval \(\Delta t_k\) is
\(\exp(-\lambda_k \Delta t_k)\). Hence, the probability that interception
occurs during step \(k\), conditioned on survival up to the start of that step,
is
\begin{equation}
    h_k = 1 - \exp(-\lambda_k \Delta t_k)
        = 1 - \exp\!\left(-\alpha\, p_{\mathrm{RR}}\bigl(\vect{p}(t_k)\bigr)\, \Delta t_k\right).
    \label{eq:hazard}
\end{equation}
This implies that longer exposure to regions with higher reachability
probability yields a larger conditional interception probability, while
preserving the bound \(0 \le h_k \le 1\).

Let \(S_{k-1}\) denote the probability that the evader survives through step
\(k-1\). Since \(h_j\) is defined as the conditional probability of
interception during step \(j\), given survival up to the start of that step,
the corresponding conditional survival probability is \(1-h_j\). Therefore, by
the chain rule of probability,
\begin{equation}
    S_{k-1} = \prod_{j=1}^{k-1} (1-h_j).
\end{equation}
The probability that interception occurs for the first time during step \(k\) is
then
\begin{equation}
    w_k = S_{k-1} h_k,
    \label{eq:prob_first_intercept}
\end{equation}
which defines the discrete first-interception distribution along the
trajectory. The total probability of at least one interception event is
\begin{equation}
    P_{\mathrm{event}} = \sum_{k=1}^{K} w_k
    = 1 - \prod_{k=1}^{K} (1-h_k).
\end{equation}

This hazard-based construction defines a discrete first-interception
distribution along the trajectory that can be evaluated efficiently within the
trajectory optimization routine. The resulting weights induce a natural
distribution over possible interception locations, which is used to compute the
expected information gain along the trajectory and to define the
sacrificial-trajectory objective.

\subsubsection{Information-Driven Trajectory Objective}

Trajectory selection is formulated as the maximization of expected geometric
information obtained from a potential interception event. Because the value of
a sacrificial trajectory changes once interception data are observed, the
objective naturally separates into pre- and post-observation regimes,
corresponding to planning under uncertainty and constraint refinement after
measurement.

\paragraph{Pre-Observation Regime.}

In the absence of interception data, the feasible launch region cannot yet
be contracted, rendering geometric information gain undefined. The primary
objective in this regime is therefore to induce an interception with high
probability so that informative measurements become available.

Assume a rectangular region of interest $\set{P}_{\textit{rect}}$ that
contains the unknown pursuer launch location. Because interception is
possible only when the sacrificial trajectory intersects the pursuer's
RR, maximizing interception probability is equivalent to
maximizing the portion of the launch region from which the pursuer could
successfully engage the trajectory.

The launch-coverage fraction induced by a trajectory $\mathcal{C}$ is
defined as
\begin{equation}
J_0(\mathcal{C})
=
\frac{
\mathrm{Area}\!\left(
\set{P}_{\textit{rect}}
\,\cap\,
\bigcup_{k=1}^{K} \set{D}\!\big(\vect{p}(t_k),\, R+r\big)
\right)
}{
\mathrm{Area}\!\left(\set{P}_{\textit{rect}}\right)
},
\label{eq:coverage}
\end{equation}
where $\mathcal{C}$ denotes the control points defining the trajectory
$\vect{p}(t)$.

\paragraph{Post-Observation Regime.}
Once interception data are observed, each event contracts the feasible
pursuer launch region by imposing a disc constraint centered at the
interception location. Let $A_{\mathrm{old}}$ denote the area of the
current feasible region, and let $A_{\mathrm{new}}(\vect{p}(t_k))$ denote
the area obtained after intersecting this region with the disc associated
with an interception at $\vect{p}(t_k)$. The resulting area reduction is
\begin{equation}
\Delta A(\vect{p}(t_k))
=
A_{\mathrm{old}} - A_{\mathrm{new}}(\vect{p}(t_k)).
\end{equation}

Trajectory selection seeks to maximize the expected contraction
of the feasible launch-region area under the first-hit distribution,
\begin{equation}
J_1(\mathcal{C})
=
\mathbb{E}[\Delta A]
=
\sum_{k=1}^{K} w_k \, \Delta A(\vect{p}(t_k)).
\label{eq:reduce_area}
\end{equation}
This objective prioritizes trajectories that are both likely to be
intercepted and expected to induce large feasible-set contractions, accelerating geometric inference of the pursuer parameters.

\paragraph{Regime-Switching Objective.}
The trajectory design objective depends on whether interception data
have been observed. Let $N_{\mathrm{int}}$ denote the number of
interception events collected thus far. The sacrificial trajectory
objective is defined as
\begin{equation}\label{eq:sac_objective}
J(\mathcal{C}) =
\begin{cases}
J_0(\mathcal{C}),
& N_{\mathrm{int}} = 0, \\[6pt]
J_1(\mathcal{C}),
& N_{\mathrm{int}} \ge 1.
\end{cases}
\end{equation}

When no interceptions have been observed, geometric contraction of the
feasible launch region is not yet possible. The objective therefore
prioritizes maximizing launch-region coverage in order to induce an
informative interception event.
Once at least one interception has occurred, trajectory design seeks to maximize the expected contraction of the feasible launch region.

\subsubsection{Trajectory Optimization}
As in Section~\ref{sec:safe_paths}, sacrificial trajectories are
parameterized using B-splines and optimized subject to kinematic feasibility
constraints that ensure flyability. Because sacrificial agents are
expendable, trajectory optimality is evaluated solely through the
information generated by a potential interception rather than travel
efficiency.

Instead of minimizing flight time, each sacrificial agent is assigned a
fixed range $R_s$. Under constant speed, this implies a fixed time of
flight
\begin{equation}
t_{f,s} = \frac{R_s}{v_s},
\end{equation}
where $v_s$ denotes the agent speed.

The spline is defined over the interval $[t_0, t_{f,s}]$ using a uniform,
unclamped knot vector
\begin{equation}
\vect{t}_k =
\big(t_0 - k\Delta_t, \ldots, t_0 - \Delta_t,\,
t_0,\,
t_0 + \Delta_t, \ldots, t_{f,s},\,
t_{f,s} + \Delta_t, \ldots, t_{f,s} + k\Delta_t\big),
\end{equation}
yielding a $C^{k-1}$-continuous trajectory whose path length is fixed by
the prescribed range $R_s$.

The resulting trajectory-design problem is posed as the constrained
optimization
\begin{subequations}
\begin{align}
\mathcal{C}_{\mathrm{opt}}
&= \arg\max_{\mathcal{C}} \; J(\mathcal{C}), \\
\text{s.t.}\quad
&\vect{p}(0) = \vect{x}_0, \\
&v_s(t_s) = v_s, \qquad \forall t_s \in \mathcal{T}_s, \\
&u_{\mathrm{lb}} \le u_s(t_s) \le u_{\mathrm{ub}}, \qquad \forall t_s \in \mathcal{T}_s, \\
&|\kappa_s(t_s)| \le \kappa_{\mathrm{ub}}, \qquad \forall t_s \in \mathcal{T}_s,
\end{align}
\end{subequations}
where $J(\mathcal{C})$ is the sacrificial objective function (Equation~\eqref{eq:sac_objective}). The kinematic feasibility constraints are
defined in Equation~\eqref{eq:kinematic}.

Taken together, the hazard model, information-driven objective, and
constrained optimization framework recast sacrificial trajectories as an
active sensing strategy that systematically contracts launch-region
uncertainty.

\section{Results}\label{sec:results}
\subsection{Simulation Procedure and Engagement-Doctrine Sensitivity}
\label{sec:results_sim_model}

This section describes the simulation framework used to generate
interception events for evaluating the proposed geometric inference and
construction pipeline. The primary objective of these experiments is to
validate the geometric methodology under the same maximum-capability
assumptions used to derive the EZ and associated reachability
constraints.

To preserve consistency with this abstraction, the pursuer is modeled as
capable of achieving its theoretical reachability limits rather than
following a specific guidance law such as proportional navigation.
Incorporating a particular guidance strategy would introduce additional
parameters—including navigation gains, maneuver limits, update rates,
autopilot dynamics, and sensing delays—that are extrinsic to the EZ
formulation and would obscure interpretation of the geometric results.

The simulation is therefore intended solely as a controlled mechanism for
generating interception observations, enabling the geometric inference
pipeline to be evaluated in isolation from guidance-law artifacts.

\subsection{Event-Based Interception Procedure}
\label{sec:event_based_intercept}

The pursuer trajectory is not explicitly modeled. Instead, interception is
determined through a geometric reachability condition consistent with a
maximum-capability pursuer.
The pursuer is parameterized by a launch location $\vect{x}_P$,
speed $v_P$, maximum travel distance $R$, and capture radius $r$, while the
evader follows a prescribed trajectory $\vect{p}_E(t)$. To represent the distance
expended by the pursuer prior to interception, a latent \emph{commitment
distance} $D \in [R_{\min}, R]$ is introduced. This parameter encodes a family of pursuer strategies: rather than engaging any evader within its RR, the pursuer may defer interception until the evader approaches a region of higher strategic value, such as a defended asset. For each trial, $D$ is drawn
from a scaled Beta distribution,
\begin{equation}
U \sim \mathrm{Beta}(\alpha,\beta), \qquad
D = R_{\min} + (R - R_{\min})U,
\end{equation}
modeling a continuum of pursuer commitment behaviors. Larger
values of $\alpha$ concentrate probability near $R$, corresponding to
late or near–maximum-range commitments, whereas larger values of $\beta$
favor shorter commitments.

Given a sampled commitment distance, interception is declared at the first
time the evader enters the capture region induced by the pursuer’s
RR,
\begin{equation}
t_{\mathrm{int}} := \inf \left\{ t \ge \frac{D}{v_P} :
\|x_E(t) - x_P\| \le D + r \right\}.
\end{equation}
If such a time exists, the interception location is recorded as
\begin{equation}
x_{\mathrm{int}} = x_E(t_{\mathrm{int}}),
\end{equation}
otherwise the trial is labeled a non-interception.

\begin{figure}[H] \centering \includegraphics[]{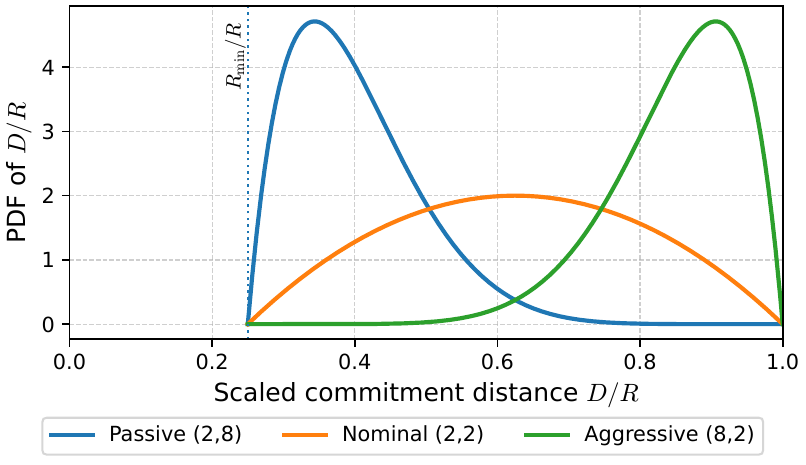} \caption{Representative commitment-distance distributions used to model variability in pursuer engagement doctrine.} \label{fig:beta} \end{figure}

Three commitment doctrines are considered, shown in
Fig.~\ref{fig:beta}: aggressive $\mathrm{Beta}(8,2)$ (green), nominal
$\mathrm{Beta}(2,2)$ (orange), and passive $\mathrm{Beta}(2,8)$ (blue).
Together, these distributions span early-, mid-, and late-commitment
behaviors, capturing a range of plausible pursuer strategies without
explicitly modeling pursuer trajectories.

Because this formulation depends only on geometric reachability rather than a
specific pursuer path, the resulting observations remain fully consistent with
the capability-based inference framework developed in
Section~\ref{sec:ez_from_interceptions}. We next evaluate the geometric inference
pipeline under this simulation procedure.

\subsection{Sequential Launch Region Refinement and Safe Replanning}\label{sec:sample_run}
\begin{figure}[h]
    \centering
    \includegraphics[width=.85\textwidth]{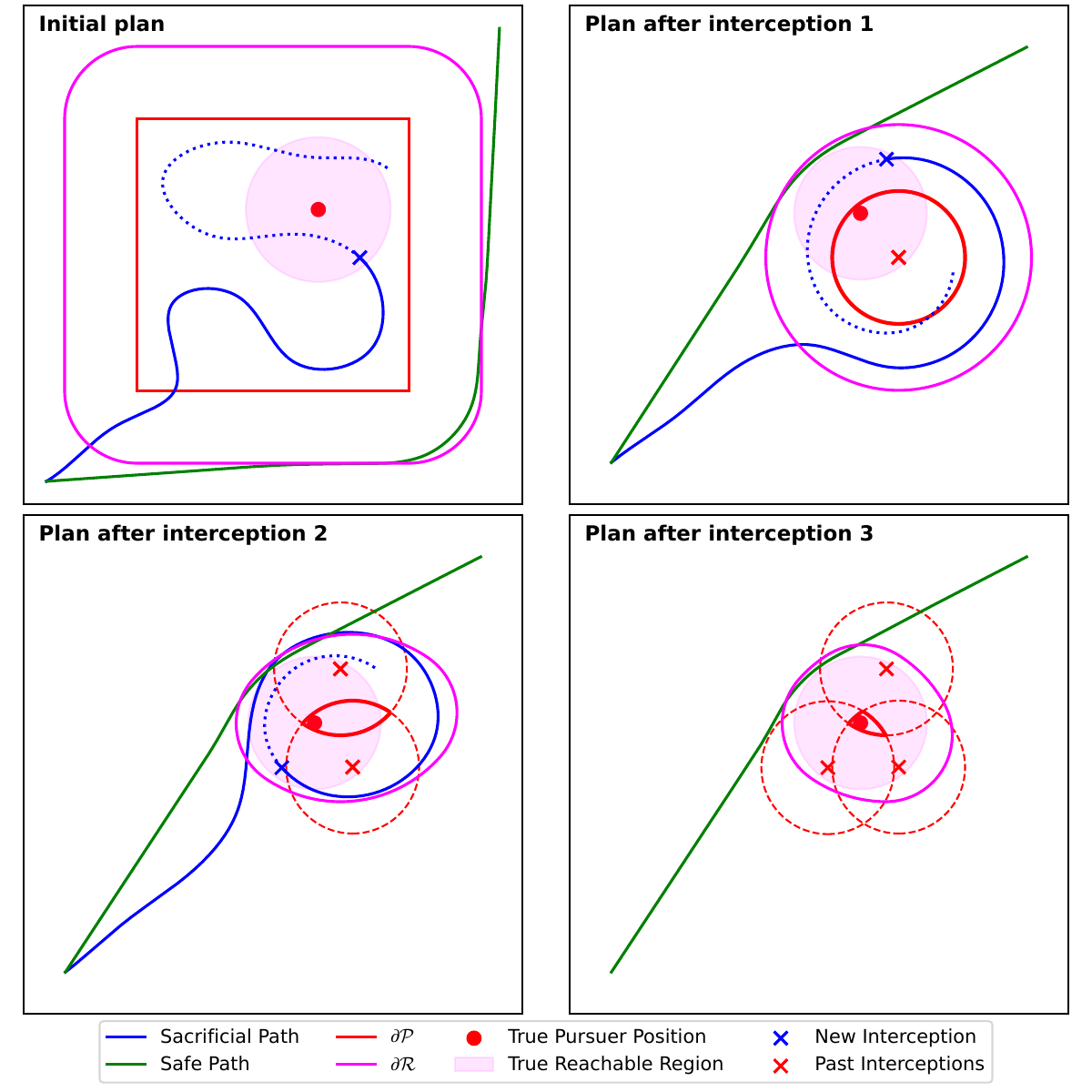}
    \caption{Sample run showing three sacrificial trajectories (blue) and four safe trajectories (green) planned with progressively increasing interception information. The feasible launch region is shown in red, RRs in magenta, and the pursuer’s true origin is marked.}
    \label{fig:sample_run}
\end{figure}

Figure~\ref{fig:sample_run} illustrates the sequential flow of the simulation.
A pursuer is assumed to originate from a known rectangular region $\mathcal{P}_{\textrm{rect}}$. Using this feasible launch region, an initial safe trajectory is generated with the path planner described in Section~\ref{sec:safe_paths}. Because the launch region is large, the resulting trajectory is conservative and relatively long, as it must avoid all EZs induced by the feasible pursuer launch region.

The corresponding RR is shown in magenta throughout the figure. The safe trajectory may enter the RR because the safety constraint is imposed on the EZ, which is a shifted version of the RR. This initial safe path is shown in green.

To reduce path length and improve operational awareness, sacrificial agents are deployed sequentially to actively refine the feasible launch region. The first sacrificial trajectory is generated using the Case~0 coverage objective described in Equation~\eqref{eq:coverage}, producing a path that provides broad coverage of the feasible launch region. After deployment, the interception location is determined using the method described in Section~\ref{sec:event_based_intercept}. Interception events are marked with red X's.

Each interception constrains the feasible launch region, shown in red in the subsequent panels. With a single interception, the constrained region becomes circular, inducing a corresponding RR shown in magenta. Using this updated operational information, a new safe trajectory is planned. This trajectory is significantly shorter than the initial path due to the reduced uncertainty in the pursuer's location.

Additional sacrificial agents can be deployed to further refine the launch region. When prior interceptions exist, sacrificial trajectories are generated using the Case~1 area-reduction objective in Equation~\eqref{eq:reduce_area}. This objective produces inward spiraling trajectories designed to place interceptions in locations that maximally contract the feasible launch region. The progressive contraction resulting from up to three interception events is shown in the figure.

The final panel presents the safe trajectory generated after all three interceptions. As the feasible launch region contracts, the safe path becomes increasingly direct, approaching the trajectory that would be obtained with perfect knowledge of the pursuer's location. The following section provides Monte Carlo results that quantitatively demonstrate this behavior across randomized trials.

\subsection{Statistical Evaluation via Monte Carlo Simulation}
Results are evaluated using 1000 Monte Carlo (MC) simulations in which the pursuer launch position is drawn uniformly from the initial rectangular feasible launch region $\mathcal{P}_{\textrm{rect}}$. Three pursuer commitment doctrines are considered, as described in Section~\ref{sec:event_based_intercept}: a passive strategy $\mathrm{Beta}(2,8)$, a nominal strategy $\mathrm{Beta}(2,2)$, and an aggressive strategy $\mathrm{Beta}(8,2)$.

For each doctrine, four evader sensing and planning configurations are evaluated, distinguished in the figures by color and line style:

\begin{enumerate}
    \item Sacrificial agents follow naive straight-line trajectories without measuring the pursuer launch time (solid blue).
    \item Sacrificial agents follow straight-line trajectories while measuring launch time (dashed blue).
    \item Sacrificial agents follow optimized spline trajectories without measuring launch time (solid green).
    \item Sacrificial agents follow optimized spline trajectories while measuring launch time (dashed green).
\end{enumerate}

Naive straight-line trajectories are generated by directing each agent toward the centroid of the current feasible launch region. If a trajectory does not result in interception, the subsequent trajectory is directed toward the boundary of the feasible region to improve geometric informativeness.

Optimized spline trajectories are generated using the objectives and methods described in Section~\ref{sec:sacrificial_path}.

\begin{figure}[H]
    \centering
    \includegraphics[width=\textwidth]{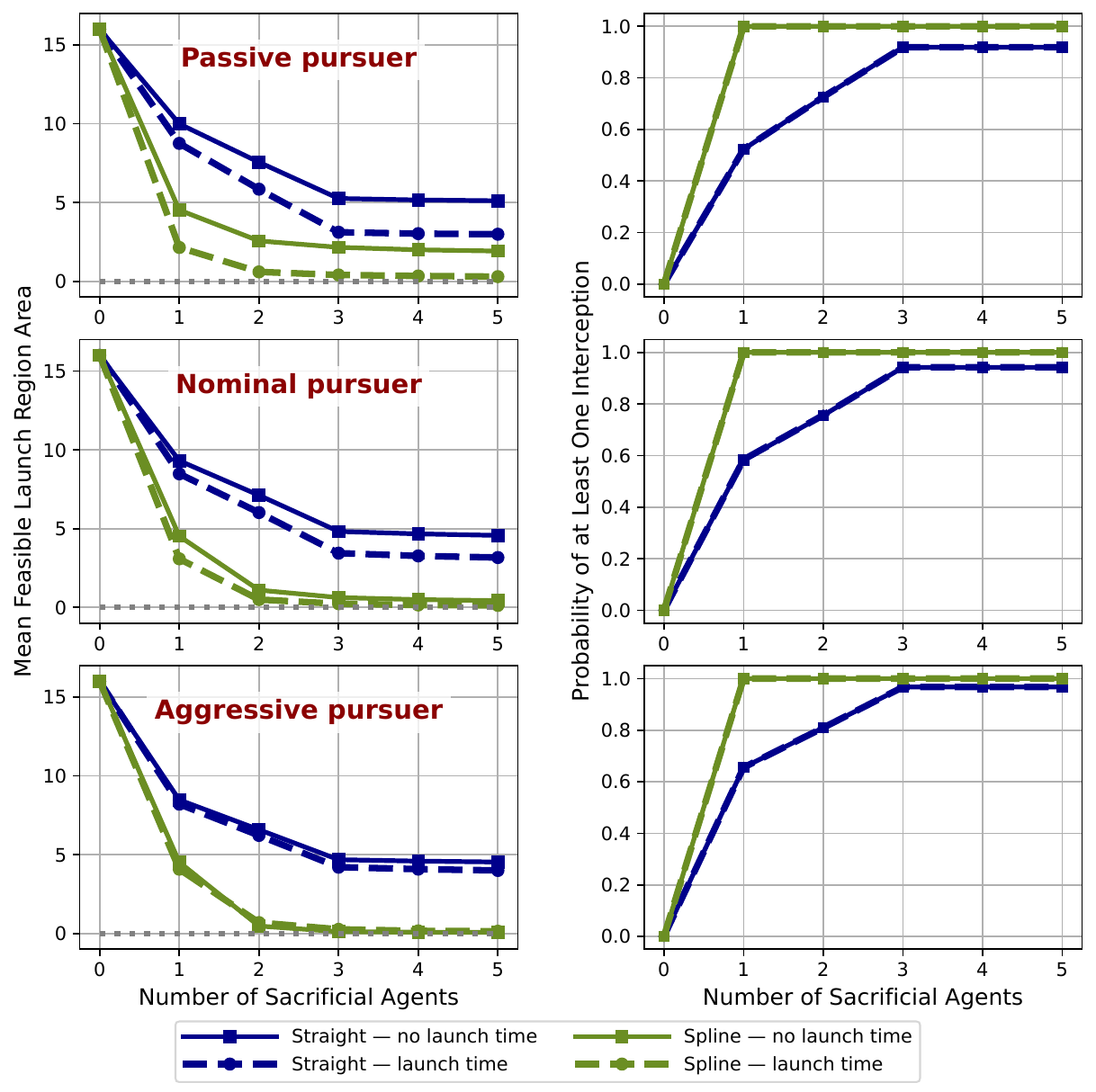}
    \caption{Mean launch region area and interception probability versus number of sacrificial agents for straight and optimized spline trajectories for passive $\mathrm{Beta}(2,8)$ (top), nominal $\mathrm{Beta}(2,2)$ (middle), and aggressive $\mathrm{Beta}(8,2)$ (bottom) commitments.}
    \label{fig:areaReduction}
\end{figure}

Each Monte Carlo trial implements the same sequential sacrificial dispatch and safe replanning procedure illustrated in Section~\ref{sec:sample_run}.

We first quantify how the feasible pursuer launch region contracts as sacrificial agents are sequentially deployed. Figure~\ref{fig:areaReduction} reports the mean planar area of the feasible launch region across $N=1000$ Monte Carlo trials for each sensing and planning configuration and for each pursuer commitment doctrine.

Across all doctrines, the optimized spline trajectories (green) reduce the feasible launch region substantially faster than the naive straight-line trajectories (blue). This behavior is most pronounced for the aggressive pursuer, where early interceptions rapidly eliminate infeasible launch locations and produce near-complete region contraction within the first few dispatches.
For the aggressive pursuer without launch-time measurement, the mean feasible launch-region area contracts from \(16.0\) initially to \(4.52\) after one sacrificial deployment, \(0.475\) after two, and \(0.120\) after three, corresponding to a \(99.25\%\) reduction by the third deployment.

The right column reports the probability that at least one sacrificial agent has been intercepted by a given dispatch. For the spline-based planner, this probability approaches unity after the first deployment across all doctrines, explaining the accelerated contraction observed in the left column. In contrast, straight-line trajectories often require multiple dispatches before an interception occurs, delaying information acquisition.

The dashed and dotted curves distinguish whether the pursuer's launch time is measured in addition to the interception location. For the aggressive pursuer, measuring launch time provides minimal additional benefit. Because the pursuer travels close to its maximum range to intercept, both the fixed-radius disks ($R+r$) and the time-dependent RRs are similar in size, resulting in comparable feasible-region constraints.

However, this effect diminishes for the nominal and passive doctrines. When launch time is not measured, each interception induces a disk of radius $R+r$ whose centers are often closely spaced, yielding a large intersection region and slower contraction. Measuring launch time instead produces tighter, path-dependent constraints that more effectively localize the pursuer. This difference is most evident for the passive doctrine, where launch-time sensing leads to significantly faster area reduction.

These results highlight an important operational implication: a pursuer that intercepts near its maximum range is inherently easier to localize, whereas intercept strategies that conserve range can help obscure the launch location when launch time is not observable.

\begin{figure}[h]
    \centering
    \includegraphics[width=\textwidth]{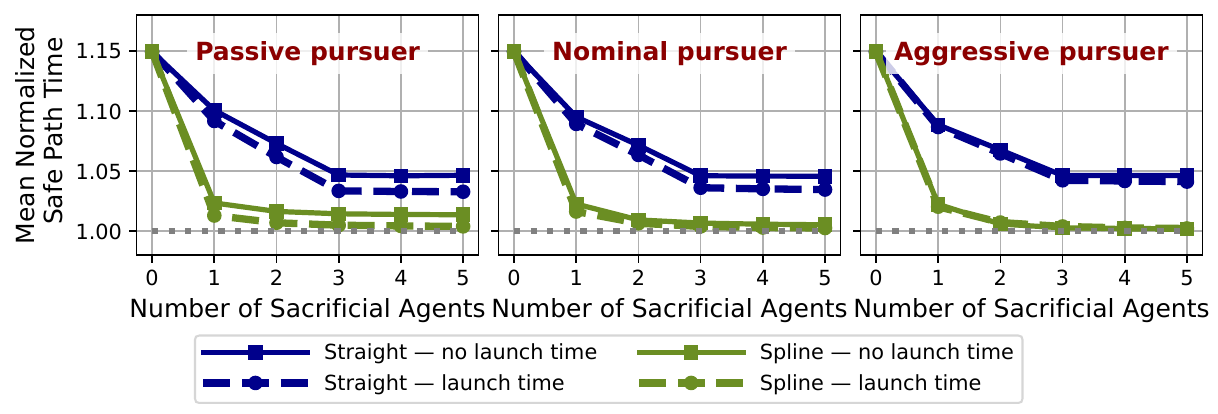}
    \caption{Mean safe path time versus number of sacrificial agents for straight and optimized spline trajectories for passive $\mathrm{Beta}(2,8)$ (left), nominal $\mathrm{Beta}(2,2)$ (middle), and aggressive $\mathrm{Beta}(8,2)$ (right) commitments.}
    \label{fig:pathLength}
\end{figure}

We next evaluate how sequential launch-region refinement translates into
operational efficiency by examining the length of the resulting safe
trajectories. To quantify this effect, we report the \emph{normalized path
length}, defined as the ratio between the safe-path length (equivalently,
travel time under constant velocity) computed using the feasible launch region
$\mathcal{P}$ after each sacrificial agent and the true optimal safe-path
time obtained under perfect knowledge of the pursuer parameters (avoiding the BEZ generated from the pursuer's actual position). A
normalized value of $1$ indicates that the sacrificial agents have
localized the pursuer sufficiently for the planner to recover the optimal
safe trajectory.

Figure~\ref{fig:pathLength} shows the mean normalized safe-path length over the $N=1000$ Monte Carlo trials for each sensing and planning configuration and for each pursuer commitment doctrine.

Across all doctrines, safe-path length decreases as additional sacrificial agents are deployed. This trend reflects the progressive contraction of the feasible launch region, which permits less conservative routing and enables trajectories that increasingly approach the true optimal safe path obtained under perfect information. The horizontal gray line in Fig.~\ref{fig:pathLength} marks a normalized value of 1, indicating recovery of the true optimal trajectory.

Consistent with the area-reduction results, spline-generated sacrificial trajectories produce faster decreases in path length than naive straight-line strategies. Early interceptions rapidly eliminate infeasible launch locations, allowing the planner to relax avoidance constraints and generate shorter paths after only a small number of dispatches. For example, for the aggressive pursuer without launch-time measurement, the mean normalized safe-path time ratio decreases from \(1.1495\) initially to \(1.0225\) after one sacrificial deployment, \(1.0061\) after two, and \(1.0029\) after three, bringing performance to within \(0.29\%\) of the perfect-information optimum by the third deployment.

The benefit of launch-time sensing again depends on the pursuer commitment doctrine. For the aggressive pursuer, measuring launch time yields only modest improvements because interceptions typically occur near the pursuer's maximum range, making the induced constraints similar with or without timing information. In contrast, the passive doctrine shows a larger separation between sensing configurations, as tighter launch-region constraints enable substantially shorter safe trajectories.

Notably, the marginal reduction in path length diminishes after several interceptions, indicating that the planned safe trajectories converge toward the true optimal path. This behavior suggests that only a small number of well-placed sacrificial deployments are required to recover most of the achievable operational benefit.

\section{Conclusion}\label{sec:conclusion}
This paper demonstrates how a relatively small number of sacrificial agents can substantially increase operational awareness and improve safety in
the presence of an unknown pursuer operating within a region of interest.

We first established a geometric framework that maps bounded uncertainty in the
pursuer launch location into bounded RRs and corresponding
EZs. This set-based construction is particularly well matched to
interception-driven inference: interception events naturally induce bounded
constraints on the unknown launch location, and these constraints propagate forward
through reachability to yield EZs with an explicit worst-case
safety interpretation.

We then presented a method for constraining the feasible pursuer launch region
directly from interception observations, including an extension that leverages
launch-time measurements to tighten the induced constraints. The resulting
feasible region defines a conservative RR and EZ that
can be enforced as a hard constraint for safe trajectory generation. In
addition, we derived a probabilistic engagement-zone construction that replaces
binary avoidance with a risk threshold, enabling a principled tradeoff between
conservatism and path length when some probability of engagement can be
tolerated.

To actively accelerate localization, we also introduced an information-driven
planner for sacrificial trajectories. When no prior interceptions are available,
the problem reduces to a coverage objective that maximizes the likelihood of
obtaining an initial interception. Once interception data exist, sacrificial
paths are selected to induce interceptions at locations that maximally contract
the feasible launch region, thereby producing rapid reductions in uncertainty.

Monte Carlo experiments validate the end-to-end pipeline. The proposed
sacrificial planner, coupled with bounded geometric engagement constraints,
achieves large improvements in launch-region contraction and safe-path
optimality with only a few sacrificial deployments, across a range of pursuer
commitment doctrines.

Several extensions are natural directions for future work. First, parallel and
cooperative dispatch of multiple sacrificial agents could replace the
sequential strategy considered here, enabling faster information acquisition
under time constraints. Second, more realistic engagement models could be
incorporated, including three-dimensional geometry, additional sensing
limitations, and alternative pursuer guidance laws. Third, the framework can be
generalized to account for uncertain or unknown pursuer capabilities (e.g.,
speed, range, and capture radius) and to integrate these uncertainties into the
inference and planning objectives. Finally, a key operational challenge is
handling multiple unknown pursuers (or multiple candidate launch locations) within
the region of interest; determining when all launch locations have been discovered
and reasoning about overlap or ambiguity between multiple feasible sets remains
a nontrivial and practically important extension.

\section*{Acknowledgments}
This work was supported by the NSF IUCRC Phase I: Center for Autonomous Air Mobility and Sensing (CAAMS) under Award No.~2139551.

Portions of this manuscript were refined and edited using OpenAI’s ChatGPT model (GPT-5). The tool was employed for language polishing, LaTeX and Python code assistance, and to help organize and clarify research ideas. All technical content, algorithms, and conclusions were developed, verified, and approved by the authors.

\bibliography{sample}

\end{document}